\documentclass{aa} 
\usepackage[varg]{txfonts}
\usepackage{graphicx}
\usepackage{txfonts}
\usepackage{natbib}
\bibpunct{(}{)}{;}{a}{}{,} 

\begin{document}

\title{White dwarf evolutionary sequences for low-metallicity progenitors:
The impact of third dredge-up } 

\author{Leandro G. Althaus\inst{1,2}, 
        Mar\'ia E. Camisassa\inst{1,2}, 
        Marcelo M. Miller Bertolami\inst{2,3}, 
        Alejandro H. C\'orsico\inst{1,2},  \and  
        Enrique Garc\'\i a--Berro\inst{4,5}}
\institute{Grupo de Evoluci\'on Estelar y Pulsaciones. 
           Facultad de Ciencias Astron\'omicas y Geof\'{\i}sicas, 
           Universidad Nacional de La Plata, 
           Paseo del Bosque s/n, 1900 
           La Plata, 
           Argentina
           \and
           IALP - CONICET
           \and
           Max-Planck-Institut f\"ur Astrophysik, 
           Karl-Schwarzschild Strasse 1, 
           85748 Garching, 
           Germany
           \and
           Departament de F\'\i sica Aplicada, 
           Universitat Polit\`ecnica de Catalunya, 
           c/Esteve Terrades 5, 
           08860 Castelldefels, 
           Spain
           \and
           Institute for Space Studies of Catalonia,  
           c/Gran Capita 2--4, Edif. Nexus 201, 
           08034 Barcelona,  Spain}
\date{Received ; accepted }

\abstract{White  dwarfs  are  nowadays   routinely  used  as  reliable
           cosmochronometers,   allowing  to   date  several   stellar
           populations.}
         {We  present  new  white  dwarf  evolutionary  sequences  for
           low-metallicity  progenitors.  This  is  motivated  by  the
           recent finding  that residual  H burning in  low-mass white
           dwarfs resulting  from $Z=0.0001$  progenitors is  the main
           energy source over a significant part of their evolution.}
         {White   dwarf  sequences   have  been   derived  from   full
           evolutionary calculations that take into account the entire
           history    of     progenitor    stars,     including    the
           thermally-pulsing  and  the  post-asymptotic  giant  branch
           phases.}
         {We  show  that for  progenitor  metallicities  in the  range
           $0.00003\lesssim Z  \lesssim 0.001$, and in  the absence of
           carbon  enrichment  due  to   the  occurrence  of  a  third
           dredge-up episode, the resulting H envelope of the low-mass
           white dwarfs is  thick enough to make stable  H burning the
           most  important energy  source  even  at low  luminosities.
           This has a significant impact on white dwarf cooling times.
           This result  is independent  of the adopted  mass-loss rate
           during the  thermally-pulsing and post-AGB phases,  and the
           planetary nebulae stage.}
         {We conclude that in the absence of third dredge-up episodes,
           a  significant  part of  the  evolution  of low-mass  white
           dwarfs  resulting   from  low-metallicity   progenitors  is
           dominated  by  stable  H  burning.   Our  study  opens  the
           possibility of  using the  observed white  dwarf luminosity
           function of low-metallicity  globular clusters to constrain
           the  efficiency  of third  dredge  up  episodes during  the
           thermally-pulsing    AGB     phase    of    low-metallicity
           progenitors.}
\titlerunning{White dwarf  evolutionary sequences  for low-metallicity
  progenitors}
\authorrunning{Althaus et al.}  

\maketitle


\section{Introduction}
\label{introduction}

White dwarf stars are the  most common end-point of stellar evolution.
These  old  stellar remnants  encode  valuable  information about  the
evolutionary  history  of their  progenitors,  providing  a wealth  of
information about  the physical  evolutionary processes of  stars, the
star  formation  history, and  about  the  characteristics of  various
stellar  populations.   Moreover,  their  structure  and  evolutionary
properties are now well  understood --- see \cite{2008PASP..120.1043F,
2008ARA&A..46..157W},  and   \cite{2010A&ARv..18..471A}  for  specific
reviews. Hence, white dwarf cooling  ages are currently considered one
of the best age indicators for a wide variety of Galactic populations,
including  open  and  globular clusters  \citep[see][for  some  recent
applications]{2009ApJ...693L...6W,                2010Natur.465..194G,
2011ApJ...730...35J,     2013A&A...549A.102B,    2013Natur.500...51H}.
Consequently,  the use  of white  dwarfs  as reliable  clocks to  date
stellar  populations has  led  to renewed  efforts  in computing  full
evolutionary  models for  these  stars, taking  into  account all  the
relevant sources and sinks of  energy, and the evolutionary history of
progenitor stars \citep{2010ApJ...717..183R, 2010ApJ...716.1241S}.

\begin{table*}
\caption{Basic  model  properties   for  sequences  with  $Z=0.00003$,
  0.0001, 0.0005 and 0.001.}
\centering
\begin{tabular}{lccccccc}
\hline
\hline
$M_{\rm ZAMS}\, (M_{\sun})$ &$M_{\rm WD}\, (M_{\sun})$ & $\log M_{\rm H}\, (M_{\sun})$ & $t_{\rm MS}$~(Gyr) & $t_{\rm He}$~(Gyr) & 
$t_{\rm 1TP}$~(Gyr) & $N_{\rm TP}$ & C/O\\
\hline
\multicolumn{8}{c}{$Z=0.00003$}\\
\hline
0.80  & 0.50611 & $-4.253$ & 11.828 & 12.602 & 12.701 &  3 & 16.221 \\
0.90  & 0.52007 & $-3.983$ &  7.741 &  8.312 & 8.403  &  3 & 10.618 \\
0.95  & 0.53487 & $-3.859$ &  6.388 &  6.888 & 6.978  &  2 & 31.084 \\
1.00  & 0.53967 & $-3.982$ &  5.330 &  5.771 & 5.859  &  3 & 20.180 \\
1.50  & 0.61840 & $-3.949$ &  1.304 &  1.485 & 1.623  & 12 & 23.882 \\
1.75  & 0.68029 & $-4.224$ &  0.773 &  0.903 & 1.042  & 19 & 23.647 \\
2.00  & 0.77021 & $-4.376$ &  0.504 &  0.586 & 0.711  & 25 & 12.736 \\
\hline
\multicolumn{8}{c}{$Z=0.0001$}\\
\hline
0.80  & 0.51976 & $-3.220$ & 11.863 & 12.741 & 12.844 &  1 &  0.270 \\
0.85  & 0.53512 & $-3.304$ &  9.500 & 10.269 & 10.367 &  3 &  0.301 \\
0.90  & 0.54839 & $-3.360$ &  7.694 &  8.347 &  8.440 &  3 &  0.295 \\
0.95  & 0.56145 & $-3.435$ &  6.329 &  6.901 &  6.998 &  4 &  0.324 \\
1.00  & 0.56765 & $-3.461$ &  5.269 &  5.793 &  5.886 &  4 &  0.332 \\
1.25  & 0.61940 & $-3.652$ &  2.462 &  2.793 &  2.886 &  8 &  0.325 \\
1.50  & 0.66588 & $-3.851$ &  1.276 &  1.486 &  1.580 & 12 &  0.264 \\
2.00  & 0.73821 & $-4.347$ &  0.581 &  0.605 &  0.749 & 23 & 11.477 \\
2.50  & 0.82623 & $-4.648$ &  0.342 &  0.348 &  0.421 & 35 & 13.355 \\
\hline
\multicolumn{8}{c}{$Z=0.0005$}\\
\hline
0.80  & 0.50932 & $-3.231$ & 12.008 & 13.124 & 13.242 &  1 & 0.241 \\
0.85  & 0.54164 & $-3.424$ &  9.551 & 10.520 & 10.622 &  3 & 0.296 \\
0.90  & 0.54619 & $-3.432$ &  7.801 &  8.601 &  8.706 &  3 & 0.240 \\
1.00  & 0.56634 & $-3.512$ &  5.335 &  5.955 &  6.059 &  5 & 0.267 \\
1.25  & 0.60391 & $-3.717$ &  2.392 &  2.797 &  2.898 &  7 & 0.304 \\
1.50  & 0.65406 & $-3.904$ &  1.337 &  1.539 &  1.638 & 12 & 0.298 \\
2.00  & 0.71244 & $-4.245$ &  0.610 &  0.633 &  0.799 & 20 & 3.202 \\
2.50  & 0.81197 & $-4.485$ &  0.357 &  0.364 &  0.441 & 31 & 1.289 \\
\hline
\multicolumn{8}{c}{$Z=0.001$}\\
\hline
0.85 & 0.53846 & $-3.434$ & 9.892 & 11.033 & 11.136 &  2 & 0.307 \\
1.00 & 0.55946 & $-3.561$ & 5.411 &  6.210 &  6.313 &  4 & 0.299 \\
1.25 & 0.60195 & $-3.747$ & 2.384 &  2.906 &  3.007 &  7 & 0.289 \\
1.50 & 0.63962 & $-3.924$ & 1.386 &  1.590 &  1.687 & 11 & 0.278 \\
1.75 & 0.66940 & $-4.035$ & 0.898 &  0.983 &  1.092 & 15 & 0.269 \\
2.25 & 0.75394 & $-4.329$ & 0.477 &  0.489 &  0.605 & 26 & 0.262 \\
\hline
\end{tabular}
\tablefoot{$M_{\rm  ZAMS}$: initial  mass, $M_{\rm  WD}$: white  dwarf
  mass, $\log  M_{\rm H}$: logarithm of  the mass of H  at the maximum
  effective  temperature  at  the  beginning of  the  cooling  branch,
  $t_{\rm MS}$:  age at the end  of the main sequence  (defined at the
  moment  when the  central  H abundance  becomes $10^{-6}$),  $t_{\rm
  He}$: age  at the beginning of  the core He burning,  $t_{\rm 1TP}$:
  age  at the  first termal  pulse,  $N_{\rm TP}$:  number of  thermal
  pulses, C/O: surface carbon to oxygen ratio after departure from the
  AGB, at $\log T_{\rm eff}=4$.}
\label{tabla1}
\end{table*}

It has  been generally accepted  that white dwarf evolution  is almost
insensitive to the initial metal  content of progenitor star.  This is
a consequence  of the  strong gravity  that characterizes  white dwarf
atmospheres. The effect  of a large surface gravity is  that metals in
the very outer layers of these  stars are depleted in relatively short
timescales, and  thus the cooling rates  do not depend on  the initial
metal  content  --   see,  for  instance,  \cite{2010ApJ...716.1241S}.
Accordingly, age determinations based on white dwarf cooling times are
not  largely affected  by uncertainties  in the  determination of  the
metallicity of the parent population.  However, the metallicity of the
progenitor stars may  play some role in determining  the core chemical
composition of a given white dwarf,  and thus could affect the cooling
rates.  In particular, because of difference in the initial progenitor
mass   --   that   could   result   from   the   dependence   of   the
initial-to-final-mass  relation  on   the  progenitor  metallicity  --
different  oxygen profiles  for  a  given white  dwarf  mass could  be
expected depending on the progenitor metallicity.

Nevertheless,  the impact  of  progenitor metallicity  on white  dwarf
evolution  could   be  much  more  relevant   than  hitherto  assumed.
Recently, \cite{2013ApJ...775L..22M} have computed  for the first time
white dwarf sequences resulting from  the full evolution of progenitor
stars with very  low metallicity ($Z=0.0001$). They found  that due to
the low  metallicity of  the progenitor stars,  white dwarfs  are born
with thicker H envelopes, leading to  more intense H burning shells as
compared with their solar metallicity counterparts. Specifically, they
found that  for this  metallicity and for  white dwarf  masses smaller
than $\sim 0.6\, M_{\sun}$, nuclear reactions are the main contributor
to   the    stellar   luminosity   for   luminosities    as   low   as
$\log(L/L_{\sun})\simeq -3.2$,  delaying the  white dwarf  cooling for
significant time  intervals.  This  is in  contrast with  the standard
assumption of neglecting  residual H burning made in  all the existing
calculations of white  dwarf cooling. This assumption  is usually well
justified because  stable H shell  burning is  expected to be  a minor
source  of  energy  for  stellar   luminosities  below  $\sim  100  \,
L_{\sun}$.  However, H burning  never ceases completely, and depending
on the  mass of  the white  dwarf and on  the precise  mass of  H left
during the previous evolutionary phases -- which depends critically on
metallicity,  see  \cite{1986ApJ...301..164I}  --   it  may  become  a
non-negligible energy source for white  dwarfs with H atmospheres.  As
shown by  \cite{2013ApJ...775L..22M} a correct assessment  of the role
played  by residual  H burning  during  the cooling  phase requires  a
detailed   calculation  of   the  white   dwarf  progenitor   history,
particularly  during  the  thermally-pulsing asymptotic  giant  branch
(AGB) phase.

In view  of this important  finding, here  we present new  white dwarf
evolutionary sequences  for a wide  range of metallicities  that cover
the  metal content  expected in  some metal-poor  stellar populations,
like  the  galactic  halo  or  the  Galactic  population  of  globular
clusters.  These sequences are of  interest for the cosmochronology of
old stellar  systems.  We  compute the full  evolution of  white dwarf
stars by  consistently evolving the  progenitor stars through  all the
stellar phases,  including the  ZAMS, the red  giant branch,  the core
helium flash (whenever it occurs), the stable core helium burning, the
AGB phase and the entire  thermally-pulsing and post-AGB phases.  With
the aim of determining at which metallicity residual H burning becomes
relevant for white dwarf evolution, we have considered four progenitor
metallicities: $Z=0.001$, 0.0005, 0.0001, and 0.00003.  To the best of
our  knowledge,   this  is  the   first  set  of   fully  evolutionary
calculations   of   white   dwarfs  resulting   from   low-metallicity
progenitors covering the relevant range  of initial main sequence and,
correspondingly, white  dwarf masses.  We emphasize  that our complete
evolutionary calculations of  the history of the  progenitors of white
dwarfs allowed us to have  self-consistent white dwarf initial models.
That means  that in our  sequences the  residual masses of  the H-rich
envelopes and  of the  helium shells  were obtained  from evolutionary
calculations, instead  of using typical values  and artificial initial
white dwarf models.

The paper  is organized  as follows.   In Sect.~\ref{code}  we briefly
describe  our  numerical  tools  and   the  main  ingredients  of  the
evolutionary  sequences, while  in Sect.~\ref{results}  we present  in
detail our evolutionary results. There, we also explore the dependence
of  the  final  H content  on  the  mass  loss  and the  treatment  of
overshooting  during  the  thermally-pulsing AGB  phase.  Finally,  in
Sect.~\ref{conclusions} we  summarize the main findings  of the paper,
and we elaborate our conclusions.

\section{Numerical setup and input physics}
\label{code}

The evolutionary  calculations presented in  this work have  been done
using     the    {\tt     LPCODE}     stellar    evolutionary     code
\citep{2003A&A...404..593A, 2005A&A...435..631A, 2012A&A...537A..33A}.
This is a well-tested and callibrated code that has been amply used in
the  study  of  different  aspects of  low-mass  star  evolution.   In
particular,  it  has  been  employed  to  compute  very  accurate  and
realistic white dwarf models, including the formation and evolution of
extremely  low-mass  white  dwarfs --  see  \cite{2008A&A...491..253M,
2010Natur.465..194G,     2010ApJ...717..897A,     2010ApJ...717..183R,
2011ApJ...743L..33M,  2011A&A...533A.139W,   2012MNRAS.424.2792C}  and
\cite{2013A&A...557A..19A}   and   references   therein   for   recent
applications.   {\tt LPCODE}  has  been tested  against another  white
dwarf evolutionary code, and uncertainties  in the white dwarf cooling
ages  arising  from the  different  numerical  implementations of  the
stellar   evolution   equations   were   found   to   be   below   2\%
\citep{2013A&A...555A..96S}.   A  detailed  description of  the  input
physics  and  numerical  procedures  can  be  found  in  these  works.
Briefly,  extra mixing  due to  diffusive convective  overshooting has
been considered during  the core H and He burning,  but not during the
thermally-pulsing  AGB.   The  breathing pulse  instability  occurring
towards  the end  of core  helium  burning has  been suppressed.   The
nuclear  network  accounts  explicitly  for  the  following  elements:
$^{1}$H,  $^{2}$H, $^{3}$He,  $^{4}$He, $^{7}$Li,  $^{7}$Be, $^{12}$C,
$^{13}$C, $^{14}$N, $^{15}$N,  $^{16}$O, $^{17}$O, $^{18}$O, $^{19}$F,
$^{20}$Ne and $^{22}$Ne, together with 34 thermonuclear reaction rates
for the pp-chains, CNO bi-cycle, and helium burning that are identical
to those  described in \cite{2005A&A...435..631A}, with  the exception
of the  reactions $^{12}$C$\  +\ $p$ \rightarrow  \ ^{13}$N  + $\gamma
\rightarrow    \    ^{13}$C    +    e$^+   +    \nu_{\rm    e}$    and
$^{13}$C(p,$\gamma)^{14}$N,       which      are       taken      from
\cite{1999NuPhA.656....3A}.    In    addition,   the    reacion   rate
$^{14}$N(p,$\gamma)^{15}$O was  taken from \cite{2005EPJA...25..455I}.
Radiative   and    conductive   opacities   are   taken    from   OPAL
\citep{1996ApJ...464..943I}   and   from   \cite{2007ApJ...661.1094C},
respectively.   The  equation  of   state  during  the  main  sequence
evolution is  that of OPAL for  H- and helium-rich composition,  and a
given metallicity.   Updated low-temperature molecular  opacities with
varying  C/O  ratios are  used,  which  is  relevant for  a  realistic
treatment  of progenitor  evolution during  the thermally-pulsing  AGB
phase \citep{2009A&A...508.1343W}.  For this  purpose, we have adopted
the  low-temperature   opacities  of   \cite{2005ApJ...623..585F}  and
presented  in \cite{2009A&A...508.1343W}.   In {\tt  LPCODE} molecular
opacities are computed by adopting the opacity tables with the correct
abundances  of  the  unenhanced  metals  (e.g.,  Fe)  and  C/O  ratio.
Interpolation  is   carried  out   by  means  of   separate  cuadratic
interpolations in $R=\rho/{T_6}^3$, $T$  and $X_{\rm H}$, but linearly
in $N_{\rm C}/N_{\rm O}$.

For  the  white  dwarf  regime,  we  use  the  equation  of  state  of
\cite{1979A&A....72..134M} for  the low-density regime, while  for the
high-density   regime   we  consider   the   equation   of  state   of
\cite{1994ApJ...434..641S},  which  accounts  for  all  the  important
contributions for both the liquid and solid phases.  We also take into
account the effects of element diffusion due to gravitational settling
\citep{1991A&A...241L..29I, 2008ApJ...677..473G}, chemical and thermal
diffusion of  $^1$H, $^3$He, $^4$He, $^{12}$C,  $^{13}$C, $^{14}$N and
$^{16}$O, see  \cite{2003A&A...404..593A} for details.  For  the white
dwarf regime and for effective temperatures lower than 10,000 K, outer
boundary conditions for the evolving  models are derived from non-grey
model  atmospheres  \citep{2012A&A...546A.119R}. All  relevant  energy
sources are taken  into account in the  simulations, including nuclear
burning, the release of latent heat of crystallization, as well as the
release of the gravitational  energy associated with the carbon-oxygen
phase separation induced by crystallization.  In particular, H burning
is considered down to $\log(L/L_{\sun}) \approx -4$.  The inclusion of
all these energy sources is done self-consistently and locally coupled
to  the   full  set   of  equations  of   stellar  evolution   --  see
\cite{2012A&A...537A..33A} for a detailed description.

The  full  calculation  of  the evolutionary  stages  leading  to  the
formation of the white dwarf is  mandatory for a correct evaluation of
the possible  role of residual  nuclear burning in cool  white dwarfs.
The present version of {\tt LPCODE} improves considerably that used in
\cite{2010ApJ...717..183R},   both  in   the  microphysics,   and  the
numerical treatment of stellar  equations and mesh distribution.  This
has allowed us  to follow the progenitor evolution  through the entire
thermally-pulsing AGB phase.  In fact,  in contrast with this previous
study,  in which  was  necessary to  artificially  modify the  opacity
profile close to the base of the convective envelope to circumvent the
numerical  difficulties arising  at the  end of  the thermally-pulsing
AGB, in  the present  work all  of the  progenitors have  been evolved
naturally throughout the entire thermally-pulsing AGB, that is, we have
not forced our models to abandon early the thermally-pulsing AGB phase
in order to get realistic white dwarf structures.

In computing  our new set  of white dwarf sequences,  overshooting was
disregarded during  the thermally-pulsing  AGB phase.  Our  reason for
this choice is that we want  to constrain how important can residual H
burning be for the evolution of low-mass low-metallicity white dwarfs.
As a result,  it is expected that  the mass of the H-free  core of our
sequences gradually  grows as  evolution goes  through the  TP-AGB. In
fact,  the  inclusion of  overshooting  at  all convective  boundaries
during the  TP-AGB leads to  a strong  enhancement of third  dredge up
events,   which  prevent   the  growth   of  the   hydrogen-free  core
\citep{2009ApJ...692.1013S}.   In the  absence of  overshooting, third
dredge up in low mass stars in  very feeble.  Hence, mass loss plays a
major role in determining the final mass of the H-free core at the end
of the TP-AGB evolution, and  thus the initial-final mass relation. In
this   work,   mass    loss   during   the   RGB    was   taken   from
\cite{2005ApJ...630L..73S}.  For the  AGB phase, we use  again that of
\cite{2005ApJ...630L..73S}  for  pulsation  periods  shorter  than  50
days. For  longer periods, mass  loss is taken  as the maximum  of the
rates of \cite{2005ApJ...630L..73S} and \cite{2009A&A...506.1277G} for
oxygen   rich    stars,   or   the    maximum   of   the    rates   of
\cite{2005ApJ...630L..73S}    and    \cite{1998MNRAS.293...18G}    for
carbon-rich  stars.   In  all  of  our  calculations,  mass  loss  was
suppressed after the  post-AGB remnants reach $\log T_{\rm  eff} = 4$.
This choice  does not bear  consequences for the  final mass of  the H
content that will characterize the white dwarf, see next section.

\begin{figure}
\centering
\includegraphics[clip,width=\columnwidth]{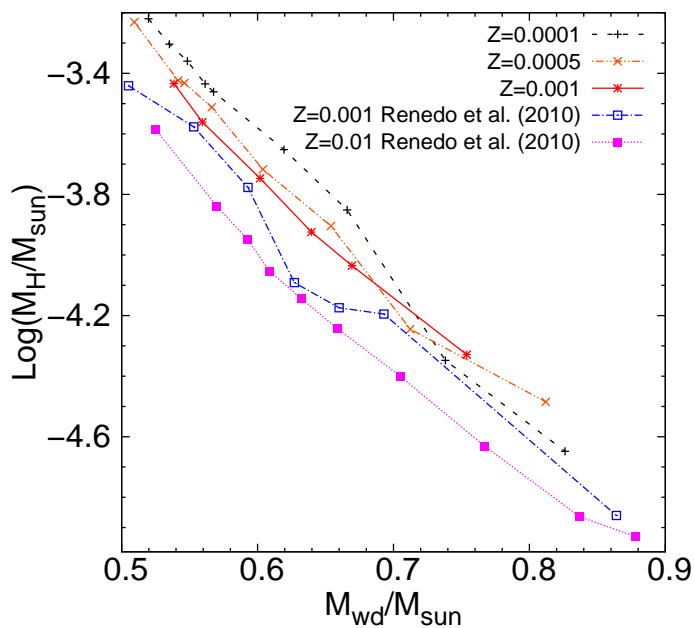}
\caption{Logarithm of  the total mass  of H  (in solar masses)  at the
  point  of  maximum effective  temperature  of  the cooling  sequence
  versus the white  dwarf mass for $Z=0.0001, 0.0005$  and $0.001$. In
  addition, we show the  predictions of \cite{2010ApJ...717..183R} for
  $Z=0.001$ and $Z=0.01$ progenitors.}
\label{Fig:envolH}
\end{figure}

The initial models for our white dwarf sequences correspond to stellar
configurations  derived from  the  full  evolutionary calculations  of
their progenitor  stars.  Four metallicities for  the progenitor stars
have  been  considered:  $Z=0.001,0.0005,0,0001$ and  $0.00003$.   The
initial  H  abundances,  by  mass, are,  respectively  0.752,  0.7535,
0.7547, and  0.7549.  A total  of 30 evolutionary sequences  have been
computed self-consistently from the ZAMS,  through the giant phase and
helium  core flash,  the  thermally-pulsing phase,  and the  planetary
nebulae stage.  They encompass a  range of initial stellar masses from
0.80 to $2.5\, M_{\sun}$. The initial  mass of the progenitor stars at
the  ZAMS and  the resulting  white  dwarf mass,  together with  other
quantities   that   will   be   discussed   below,   are   listed   in
Table~\ref{tabla1}.  In all cases, the  white dwarf evolution has been
computed  down   to  $\log(L/L_{\sun})=-5.0$.    Moreover,  additional
sequences  have been  computed to  assess the  impact on  the final  H
content due to mass loss and overshooting (and the occurrence of third
dredge-up in  low-mass progenitors)  during the  thermally-pulsing AGB
phase, see Sect. \ref{mlossov}.

\begin{figure}
\centering
\includegraphics[clip,width=\columnwidth]{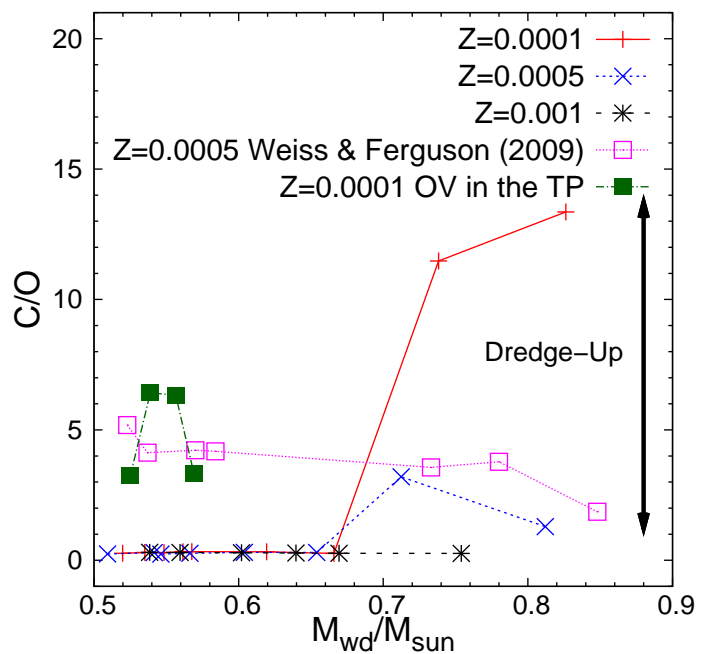}
\caption{Ratio of  the carbon to oxygen  abundances at the end  of the
  thermally-pulsing AGB  phase as a  function of the white  dwarf mass
  for $Z=0.0001,  0.0005$, and $Z=0.001$ progenitors  -- solid, dotted
  and dashed lines, respectively. In addition, we show with filled and
  hollow  squares   the  predictions  from  our   low-mass  $Z=0.0001$
  progenitors in the  case that overshooting is  considered during the
  thermally-pulsing  AGB  phase  and   the  corresponding  results  of
  \cite{2009A&A...508.1343W} for $Z=0.0005$ progenitors.}
\label{Fig:CO}
\end{figure}

\section{Evolutionary results}

\label{results}

\begin{figure*}
\centering
\includegraphics[clip,width=\textwidth]{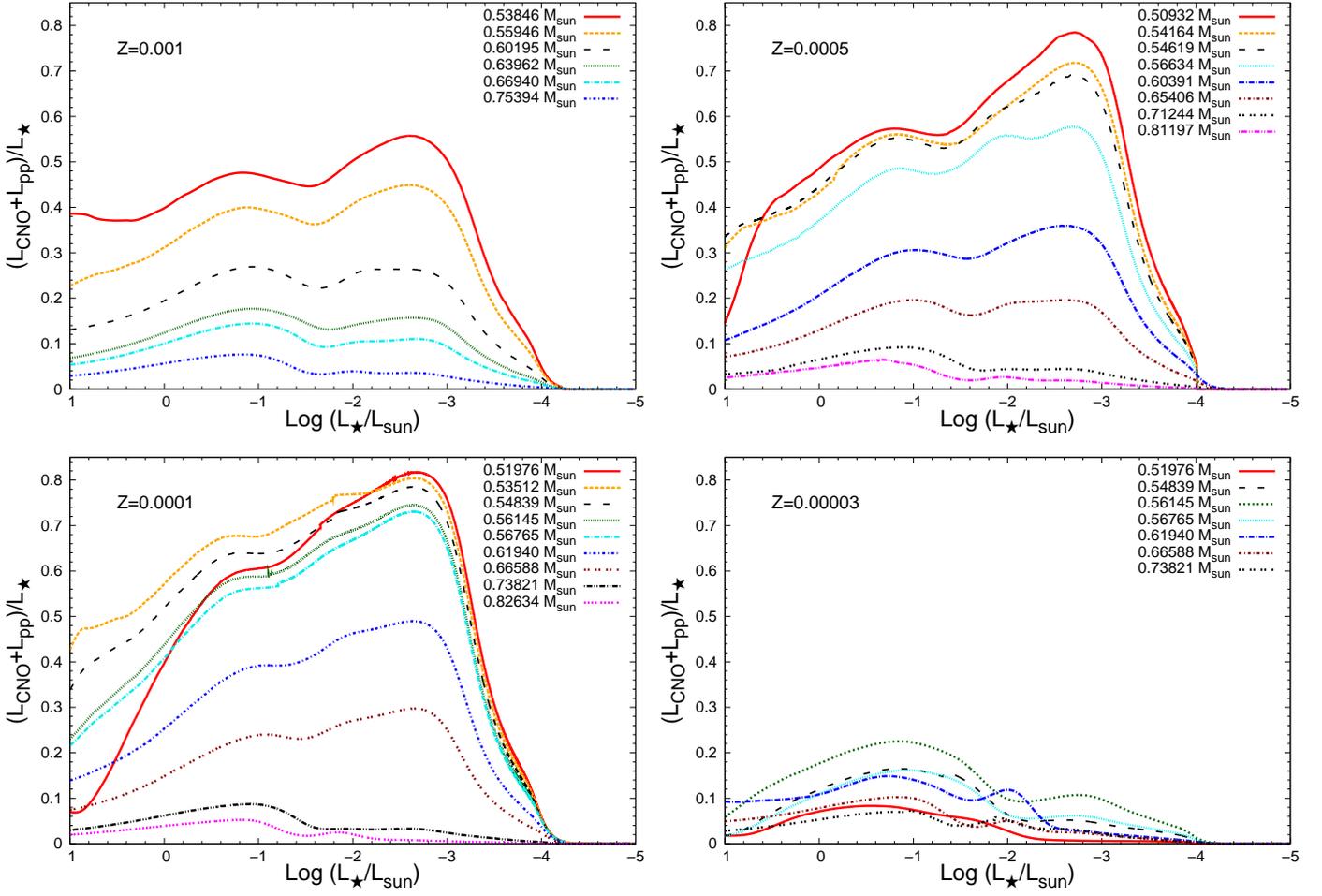}
\caption{Fraction of the total white dwarf luminosity due to H nuclear
  reactions  for  the four  different  metallicities  studied in  this
  work. Note  that for $Z=0.0001$  and $Z=0.0005$, residual  H burning
  becomes the main energy source in low-mass white dwarfs, even at low
  luminosities.}
\label{Fig:Lreacnuc}
\end{figure*}

We will see that in the absence of third dredge-up episodes during the
AGB  phase, most  of  the  evolution of  white  dwarfs resulting  from
low-mass  ($M\la  1.25  \,M_{\sun}$)  low-metallicity  progenitors  is
dominated  by stable  H burning.   In  particular, the  occurrence of  third
dredge-up reduces the final mass H content with which the white dwarfs
enter their cooling track, thus  weakening the role played by residual
H burning  (see Sect.  \ref{mlossov}).  Contrary  to the  situation of
metal-rich  progenitors, for  which  there  exist direct  observational
inferences about the  occurrence of third dredge-up in  AGB stars with
stellar masses larger  that $M\sim 1.5\,M_{\sun}$ \citep{2008A&A...486..511L}, 
the situation for their metal-poor counterparts seems  to be not clear, 
particularly for the low-mass AGB stars.

Evidence  pointing  towards  the  occurrence  of  third  dredge-up  in
progenitor stars with very low metallicity is provided by observations
of carbon-enhanced metal-poor stars  (CEMP) stars. However, the origin
of the carbon-enhancement  for CEMP stars is not  fully understood.  A
possible scenario involves binary pairings of a low-mass star (the one
observed today)  and a more  massive binary companion that  can donate
carbon in a  mass-transfer event. This requires that  the more massive
star to be in  the right mass range to pass through  an AGB phase with
significant  production of  carbon (and  s-process elements),  roughly
between 1.3 and $6.0   \,M_{\sun}$  \citep{2014ApJ...788..180C}.   Thus,  the   CEMP
phenomenon is associated with massive and intermediate-mass progenitor
stars,  and not  with the  low-mass stars  that are  relevant for  our
study.  It is worth noting that the third dredge-up evidence suggested
by the CEMP stars is in  line with our calculations for the $Z=0.0001$
progenitors, which predict  a strong carbon enrichment {\sl even} in the absence
of overshooting for stellar masses  larger than $\sim 1.5 \, M_{\sun}$
(Table~\ref{tabla1}).

This  evidence  is  not  conclusive  about  the  occurrence  of  third
dredge-up  episodes in  low-mass,  low-metallicity AGB  stars. On  the
theoretical side,  turning on convective  overshooting at the  base of
the convective envelope artificially  triggers the occurrence of third
dredge-up \citep{1999A&A...344..617M}. But 
the  initial-final mass  relation of  solar metallicity  stars suggests
that third dredge-up episodes should be less relevant than
predicted by evolutionary models that include overshooting at all convective 
boundaries during the TP-AGB phase
\citep{2009ApJ...692.1013S}.  This expectation is  in line with simple
hydrodynamical  arguments suggesting  that the penetration of  the convective envelope
  though the  H/He transition  is  expected to  be minor.  A
simple estimate can be done on the basis of the bulk Richardson number
(Ri$_{\rm B}$).  In  particular, \cite{2007ApJ...667..448M}  proposed that
the bulk Richardson number could be used as a dimensionless estimation
of the ``stiffness``  of a stable layer against  convection. Thus,
this dimensionless number could be  used  to constrain  the  strength of  extra-mixing
episodes.  These authors  suggest that high values  of Ri$_{\rm B}$ should
inhibit the  occurrence of apreciable  extra-mixing.  This is  in line
with terrestrial  simulations which show that  the entrainment process
in the  ocean is estimated  to operate  up to Ri$_{\rm B}  \sim 10^5-10^6$
\citep{2007ApJ...667..448M}.

In  order to  assess the  possibility that  the outer  convection zone
penetrates into  the carbon-rich layers,  we have  calculated Ri$_{\rm B}$
along  de  H/He transition  for  our  $0.90 M_{\odot}$  sequence  with
$Z=0.0001$,  following    Eq.~(2)  in  \cite{2007ApJ...667..448M}.
This calculation was performed at the moment when the outer convective
zone reaches its maximum depth, during the second thermal pulse, when  
third dredge-up episodes  are expected to occur.   We find that the
 Ri$_{\rm B}$ values  range from $\sim 1$ in the  H rich zone, up to $\sim 5 \times
10^5$ at the  outermost layers that has been enriched in carbon  by the pulse
driven  convection  zone.   Although  this does  not  mean  that  some
extra-mixing should not occur, this last value is high enough to prompt us to
suspect that is not entirely correct to assume  the occurrence of 
extra-mixing  through the entire H-He transition layers. Hence, 
on the basis of this hydrodynamical argument,  it cannot be 
conclusive about the occurrence of  appreciable third  dredge-up  episodes 
in  these low-mass  stars. In view of these results, we believe to be
worthwhile to compute sequences considering and disregarding extra-mixing 
during the thermally-pulsing AGB phase, and explore the consequences for
white dwarf evolution. 

In  all of  our  sequences,  evolution starts  from  the  ZAMS and  is
followed through the  stages of stable H and helium  core burning, the
stage of mass loss during the entire thermally-pulsing AGB, the domain
of the  planetary nebulae at  high effective temperature,  and finally
the  terminal  white  dwarf  cooling track,  until  very  low  surface
luminosities are reached.   For most of them,  evolution has proceeded
to the helium core  flash on the tip of the Red  Giant Branch, and the
following  recurrent sub-flashes,  before  the  progenitors reach  the
stable core helium-burning  stage on the Horizontal  Branch. The final
white dwarf mass (in solar masses)  of these sequences can be found in
Table~\ref{tabla1}, together with the corresponding evolutionary times
(in Gyr) for the main relevant stages of progenitor stars. Also listed
in this table are the total mass of residual H at the point of maximum
effective temperature at the onset of the cooling track, the number of
thermal pulses on  the AGB, and the  final C/O ratio after  the end of
the thermally-pulsing AGB phase.

An expected result shown in  Table~\ref{tabla1} is that the residual H
content left by prior evolution  decreases with increasing white dwarf
masses, a  fact that  helps to understand  the dependence  of residual
nuclear burning  on the stellar  mass discussed below.   Another trend
that can be seen in Table~\ref{tabla1}  is that the residual H content
displays   a   marked   dependence   on   the   metallicity   of   the
progenitor.   This  is   related   to  the   well-known  result   that
low-metallicity  progenitors depart  from  the AGB  with more  massive
envelopes  \citep{1986ApJ...301..164I}, leading  to white  dwarfs with
thicker H  envelopes.  Indeed, we find  that, in general, for  a given
white dwarf  mass, a larger  H content is expected  to be left  as the
metallicity  progenitor  decreases,  the   only  exception  being  the
sequences with  the lowest  metallicity ($Z=0.00003$).  This  can best
appreciated in  Fig.~\ref{Fig:envolH}, where the  total mass of  H (in
solar  units) obtained  in our  calculations at  the beginning  of the
cooling  branch is  shown  for  $Z=0.0001, 0.0005$  and  $0.001$ as  a
function  of  the   white  dwarf  stellar  mass.    In  addition,  the
predictions of  \cite{2010ApJ...717..183R} for $Z=0.001$  and $Z=0.01$
progenitors are shown.   As a result of the more  massive H envelopes,
residual H  burning is  expected to become  more relevant  in low-mass
white  dwarfs  with low-metallicity  progenitors,  see  later in  this
section.

We  will  show in  Sect.~\ref{mlossov}  that  for a  given  progenitor
metallicity, the  amount of H  with which  the remnant will  enter the
cooling  branch is  intimately connected  with the  occurrence of  the
third dredge-up  during the  thermally-pulsing AGB phase.   During the
third dredge-up, the  envelope becomes carbon-enriched as  a result of
surface  convection penetrating  into carbon-rich  deeper layers.   In
this regard in Fig.~\ref{Fig:CO} we  display the final ratio of carbon
to oxygen abundances at the end of the thermally-pulsing AGB phase, as
a function of the white dwarf  mass for $Z=0.0001$, 0.0005, and 0.001.
This figure shows that for lower metallicities the effect of the third
dredge-up  on the  C/O  ratio  is pronounced.   We  note  that in  our
calculations we  did not find  any third dredge-up episode  except for
the two more  massive sequences -- those with initial  ZAMS masses 2.0
and  $2.5\,   M_{\sun}$  --  with  $Z=0.0001$   and  $Z=0.0005$.   The
occurrence  of  third  dredge-up  in these  sequences  is  clear  from
Fig.\ref{Fig:CO}, where a  sudden increase in the final  C/O ratio can
be seen -- at a white dwarf mass of $\sim\, 0.65 M_{\sun}$.  The third
dredge-up reduces the amount of mass  of residual H, as it is inferred
from  the  change  in the  corresponding  curves  shown  in
Fig.\ref{Fig:envolH}.  Since the role played by nuclear burning during
white dwarf evolution  is strongly tied to the residual  H mass, which
is essentially a function of the occurrence of third dredge-up and the
metallicity of the progenitor, our  derived values for the H envelopes
should be considered as upper  limits.  In fact, as already mentioned,
we disregarded overshooting during the thermally-pulsing AGB phase. In
particular, in our simulations third  dredge-up is much less efficient
in  the  absence  of  overshooting.  Thus  our  choice  of  neglecting
overshooting favors the formation of  white dwarfs with larger amounts
of H (see Sect.\ref{mlossov}).

In connection  with this, we  mention that the mass  of the H  that is
left on the white dwarf  (listed in Table~\ref{tabla1}) is independent
on the occurrence of mass loss  during the planetary nebulae stage. As
mentioned before, for all our sequences mass loss was suppressed after
the post-AGB remnants reach $\log T_{\rm eff} = 4$. To test the impact
of additional mass loss on the residual mass of H, we have re-computed
the  post-AGB  evolution  of   our  $0.80\,  M_{\sun}$  sequence  with
$Z=0.0005$ from  this effective temperature, but  now considering mass
loss during all the planetary  nebulae stage, including the hot stages
of white dwarf evolution, following  the prescription of the mass loss
rates  for  hot  central  stars   of  planetary  nebulae  provided  by
\cite{1995A&A...299..755B}       from       the       results       of
\cite{1988A&A...207..123P}. We find that the  mass of H with which the
white dwarf enters its cooling track  is not altered by the occurrence
of  mass  loss during  these  stages.   Mass  loss during  this  phase
increases the rate  of evolution of the post-AGB remnant  (since the H
envelope is  reduced now by both  nuclear burning and mass  loss) with
the result that the remnant reaches a given effective temperature with
the same  mass of the H  content irrespective of whether  mass loss is
considered or not.

We remark that the residual masses of H in our $Z=0.001$ sequences are
somewhat   larger   than   those   found   in   our   previous   study
\citep{2010ApJ...717..183R}    for   the    same   metallicity,    see
Fig.~\ref{Fig:envolH}.  This might be due to the several numerical and
physical updates in {\tt LPCODE}.  In particular, in contrast with our
previous study, here  we have not artificially altered  the opacity at
the base  of the  convective envelope  in order  to get  initial white
dwarf structures (see Sect.~\ref{code}).   Instead, our sequences were
computed  self-consistently  throughout  this  stage  and  during  the
departure from the AGB.

\begin{figure}
\centering
\includegraphics[clip,width=\columnwidth]{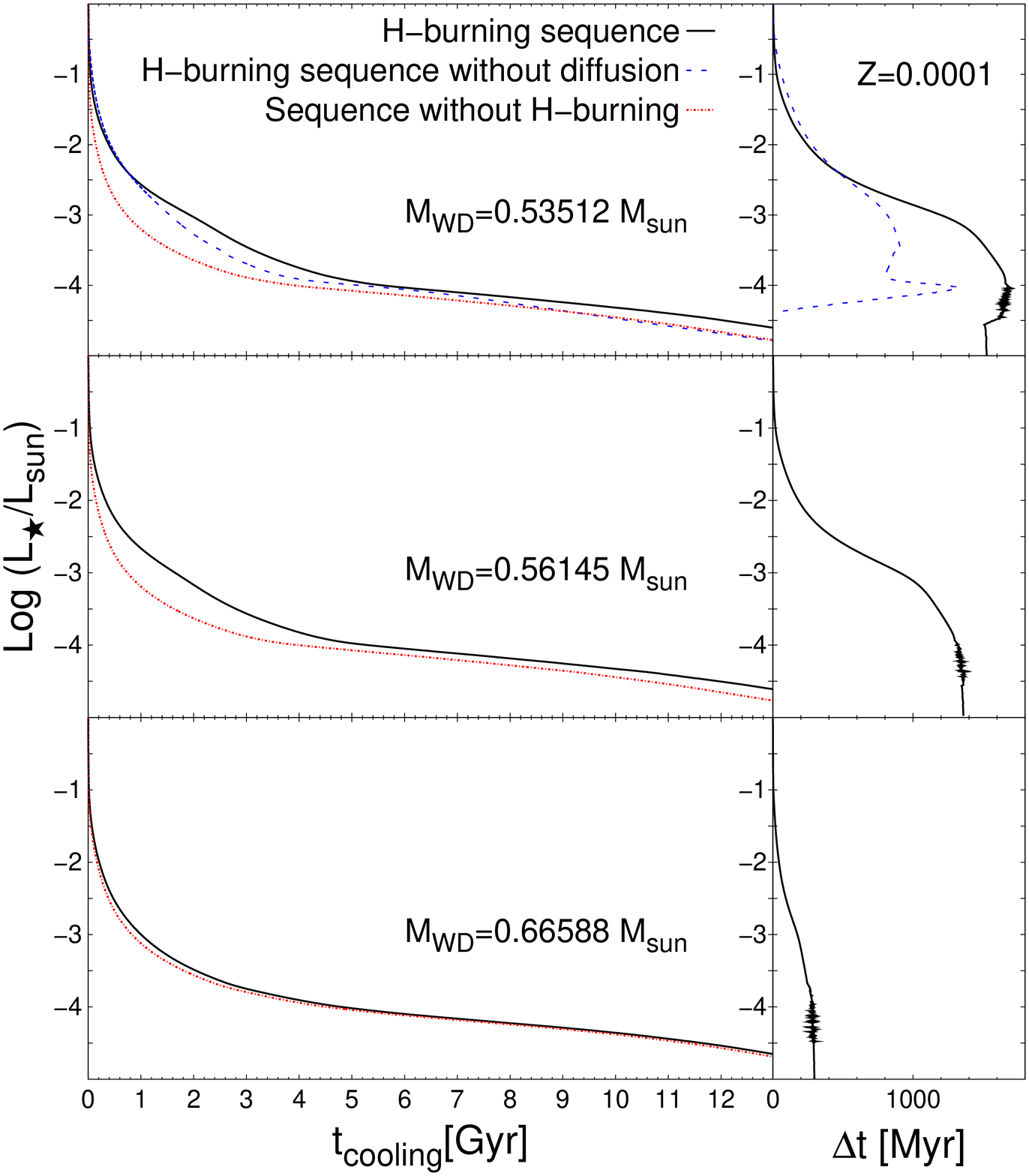}
\caption{Impact of  residual nuclear burning  on the cooling  times of
  three selected white dwarf sequences resulting from progenitors with
  $Z=0.0001$. The solid line illustrates  the evolution in the case in
  which H burning is taken into account, whereas the dotted line shows
  the case in which nuclear burning is disregarded. For the $0.53512\,
  M_{\sun}$ white  dwarf cooling sequence  we also show  the evolution
  when diffusion is neglected (dashed line).}
\label{Fig:Lvst0.0001}
\end{figure}

\begin{figure}
\centering
\includegraphics[clip,width=\columnwidth]{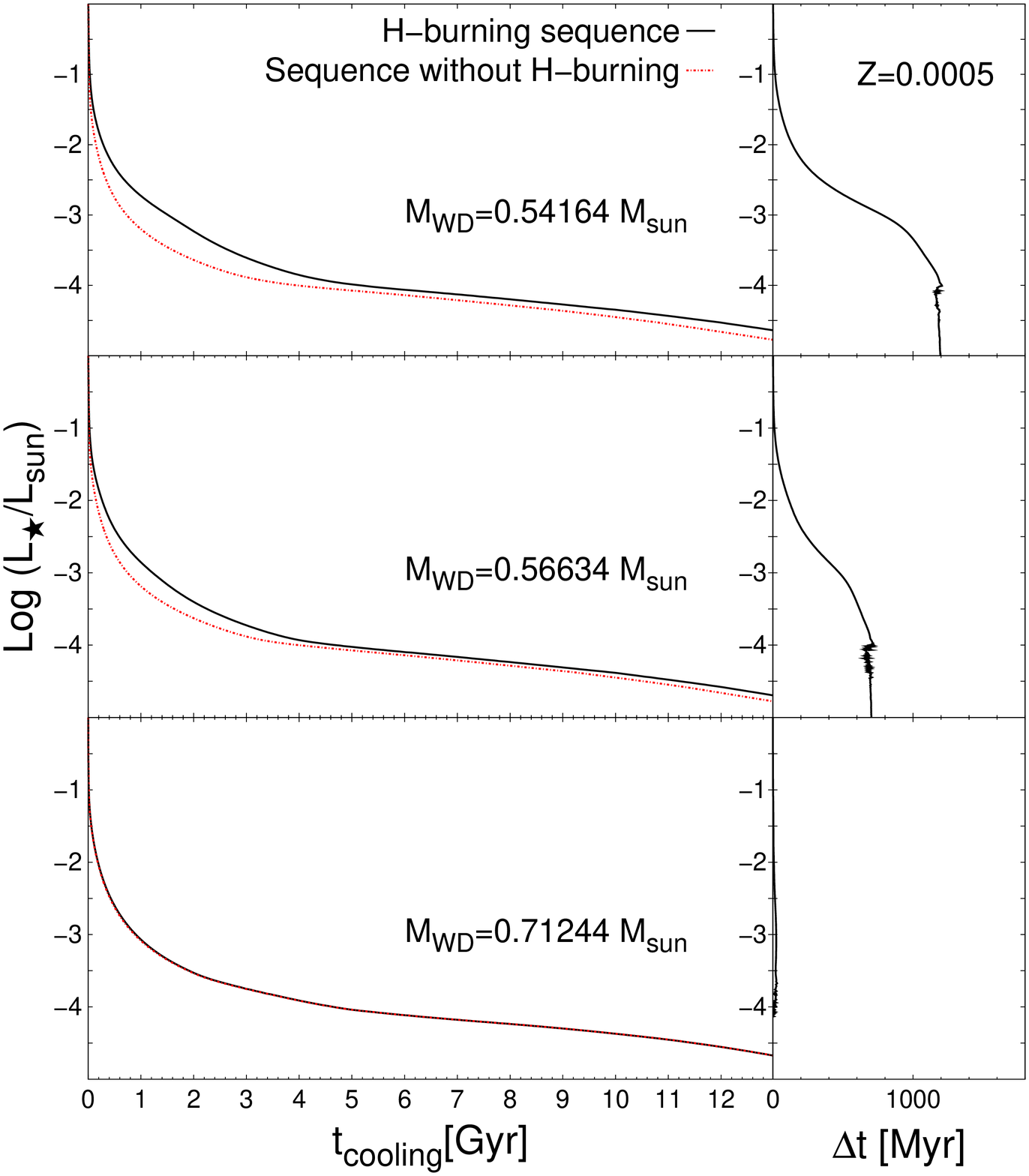}
\caption{Same  as Fig.~\ref{Fig:Lvst0.0001}  but for  the white  dwarf
  sequences of progenitors with $Z=0.0005$.}
\label{Fig:Lvst0.0005}
\end{figure}

As a  result of the  more massive H  envelopes, residual H  burning is
expected to become more relevant  in white dwarfs with low-metallicity
progenitors.  This  was borne  out by  \cite{2013ApJ...775L..22M}, who
found that for white dwarfs with progenitors of metallicity $Z=0.0001$
stable H  burning becomes one of  the main energy sources  of low-mass
white  dwarfs  for  substantial  periods of  time.   The  calculations
performed  here show  that this  is also  true even  for white  dwarfs
resulting from  progenitors with  metallicities as large  as $Z\approx
0.001$.  This is inferred  from Fig.~\ref{Fig:Lreacnuc}, where we plot
the fraction  of the total luminosity  that results from H  burning at
different  stages  of the  white  dwarf  cooling  phase for  the  four
metallicities explored in  our work.  It is clear  that the luminosity
of low-mass  white dwarfs resulting from  progenitors with metallicity
in the  range $  0.0001\la Z  \la 0.001 $  is completely  dominated by
nuclear burning,  even at rather low  luminosities. Specifically, note
that for white dwarfs with $M\la 0.6 \,M_{\sun}$ the energy release of
H  burning is  the  main energy  source  at intermediate  luminosities
($-3.2 \la  \log (L/L_{\sun})\la -1$,  or $7800 \la  T_{\rm eff}(K)\la
26,000$).  In  sharp contrast, in  the case of white  dwarfs resulting
from  progenitors with  $Z=0.00003$,  the role  played  by residual  H
burning is  much less relevant.   At such very low  metallicities, all
the progenitor  sequences we computed  experience a H  ingestion flash
during  the  first  thermal  pulses,  resulting  in  a  strong  carbon
enrichment of the outer layers  (see Table~\ref{tabla1}), and in white
dwarfs with  thin H envelopes  that are unable to  sustain appreciable
nuclear burning.   Hence, according  to our  calculations, progenitors
with metallicities lower than $Z\sim 0.00003$ are not expected to form
white dwarfs in which residual H burning is appreciable.

\begin{figure}
\centering
\includegraphics[clip,width=\columnwidth]{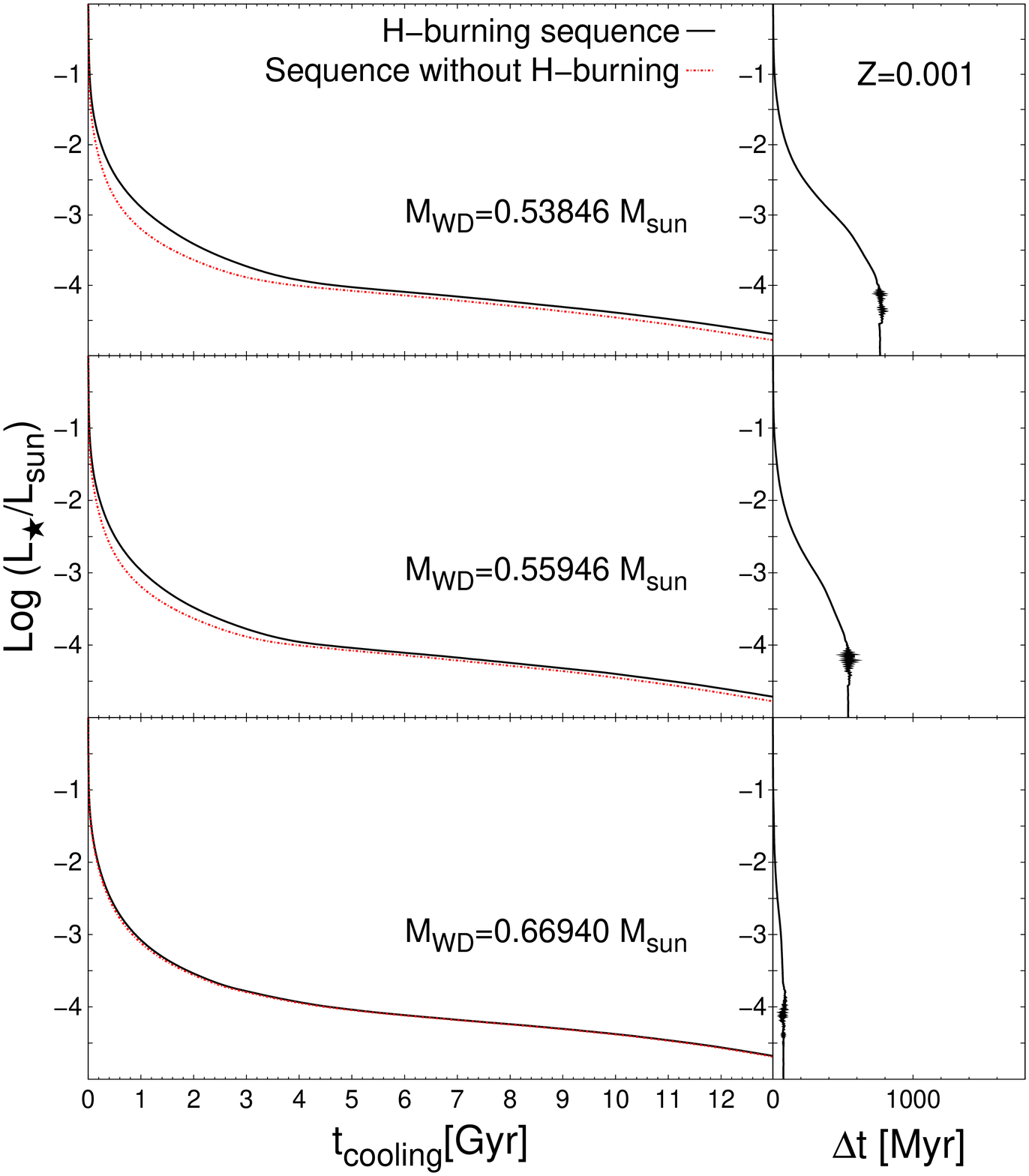}
\caption{Same as  Fig.~\ref{Fig:Lvst0.0001} but for the  case of white
  dwarf progenitors with $Z=0.001$.}
\label{Fig:Lvst0.001}
\end{figure}

The impact of  residual H burning on the cooling  times is depicted in
Figs.~\ref{Fig:Lvst0.0001},          \ref{Fig:Lvst0.0005},         and
\ref{Fig:Lvst0.001} for progenitor metallicities of $Z=0.0001$, 0.0005
and 0.001, respectively, and for three selected white dwarf sequences.
The solid lines display the cooling  times (in Gyr) of the sequence in
which H  burning is considered,  while the dotted lines  correspond to
the situation in  which residual H burning is  disregarded.  The right
panels  of  each figure  display  the  corresponding delays  (in  Myr)
introduced  by nuclear  burning.   Clearly, nuclear  burning leads  to
substantial  delays  in  the  cooling  times,  as  compared  with  the
situation  encountered   in  white  dwarfs  that   result  from  Solar
metallicity progenitors,  for which  nuclear burning  does not  play a
leading role,  and most of the  energy release comes from  the thermal
energy  stored in  the interior.   This is  most noticeable  for those
white dwarfs  that result  from progenitor  stars with  metallicity of
$Z=0.0001$ -- \cite[see][]{2013ApJ...775L..22M} --  but also for white
dwarfs   resulting   from   progenitors  with   larger   metallicities
($Z=0.0005$).  Neglecting the energy released by nuclear burning leads
to  a marked  underestimation of  the cooling  times, by  more than  a
factor    of    2    at   intermediate    luminosities.     Even    at
$\log(L/L_{\sun})\sim -4$, residual H burning  leads to an increase in
the  cooling times  of the  low-mass  white dwarfs  by $20-40\%$  (see
Table~\ref{tabladeltat}).   Note  also  that for  more  massive  white
dwarfs,  the  impact of  H  burning  on  cooling  times is  much  less
relevant.   In  the  case   of  progenitors  with  larger  metallicity
($Z=0.001$), the delay introduced by H burning is also apparent in the
least  massive  white  dwarfs,  and,  as  expected  from  the  earlier
discussion,  more relevant  than  found in  our previous  calculations
\citep{2010ApJ...717..183R} for the same  metallicity. Here, H burning
increases the cooling times  by $10-20\%$ at $\log(L/L_{\sun})\sim -4$
in the case of low-mass white  dwarfs. It is clear that stable nuclear
burning  in low-mass,  low-metallicity white  dwarfs can  constitute a
main energy source, delaying substantially  their cooling times at low
luminosities.

\begin{table*}
\caption {White dwarf  cooling ages for H burning  sequences and their
  corresponding time delays compared to the sequences in which nuclear
  burning is neglected.}  \centering
\begin{tabular}{lccccccccc}
\hline
\hline
\multicolumn{1}{c}{$-\log(L/L_{\sun})$} & \multicolumn{9}{c}{$t_{\rm cool}$~(Gyr)}\\
\hline
\multicolumn{1}{c}{} & \multicolumn{3}{c}{$Z=0.0001$} & 
\multicolumn{3}{c}{$Z=0.0005$} & \multicolumn{3}{c}{$Z=0.001$}\\
\hline
\multicolumn{1}{c|}{} & $0.535\, M_{\sun}$ & $0.561\, M_{\sun}$ & $0.666\, M_{\sun}$ & $0.542\, M_{\sun}$& $0.566\, M_{\sun}$ & $0.712\, M_{\sun}$ & $0.538\, M_{\sun}$ & $0.559\, M_{\sun}$& $0.669\, M_{\sun}$\\
\hline
1.0 &  0.04 &  0.03 &  0.02 &  0.03 &  0.03 &  0.01 &  0.03 &  0.02 &  0.01 \\
2.0 &  0.40 &  0.33 &  0.21 &  0.29 &  0.27 &  0.19 &  0.26 &  0.22 &  0.19 \\	 					
3.0 &  1.95 &  1.67 &  1.01 &  1.54 &  1.23 &  0.92 &  1.20 &  1.07 &  0.91 \\		
3.5 &  3.13 &  2.81 &  2.05 &  2.66 &  2.28 &  1.92 &  2.26 &  2.09 &  1.88 \\
4.0 &  5.61 &  5.30 &  4.81 &  5.24 &  4.78 &  4.66 &  4.68 &  4.62 &  4.58 \\
4.5 & 12.05 & 11.97 & 11.68 & 11.74 & 11.32 & 11.57 & 11.37 & 11.16 & 11.45 \\
5.0 & 16.53 & 16.09 & 15.58 & 16.19 & 15.51 & 15.28 & 15.75 & 15.38 & 15.40 \\
\hline
\multicolumn{1}{c}{} & \multicolumn{9}{c}{$\delta t$~(Gyr)}\\
\hline
1.0 & 0.03 & 0.02 & 0.00 & 0.02 & 0.01 & 0.00 & 0.01 & 0.01 & 0.00 \\
2.0 & 0.25 & 0.18 & 0.04 & 0.14 & 0.11 & 0.01 &	0.11 & 0.07 & 0.01 \\
3.0 & 1.21 & 0.92 & 0.18 & 0.79 & 0.48 & 0.02 &	0.45 & 0.31 & 0.06 \\ 
3.5 & 1.53 & 1.17 & 0.23 & 1.06 & 0.62 & 0.02 &	0.63 & 0.43 & 0.07 \\	
4.0 & 1.68 & 1.31 & 0.28 & 1.21 & 0.71 & 0.01 &	0.75 & 0.53 & 0.09 \\ 
4.5 & 1.58 & 1.36 & 0.29 & 1.18 & 0.70 & 0.00 &	0.78 & 0.54 & 0.07 \\
\hline
\end{tabular}
\label{tabladeltat}
\end{table*}

To assess  the influence of  element diffusion  on H burning,  we have
recomputed  the  evolution of  the  $0.53512\,  M_{\sun}$ white  dwarf
sequences with $Z=0.0001$ for the  extreme situation of neglecting all
of the  diffusion processes.  The  resulting cooling time is  shown in
Fig.~\ref{Fig:Lvst0.0001} as  a dashed  line. Note  that even  in this
extreme case, substantial delays in  the cooling times are expected as
a result of H burning.

\begin{figure}
\centering
\includegraphics[clip,width=\columnwidth]{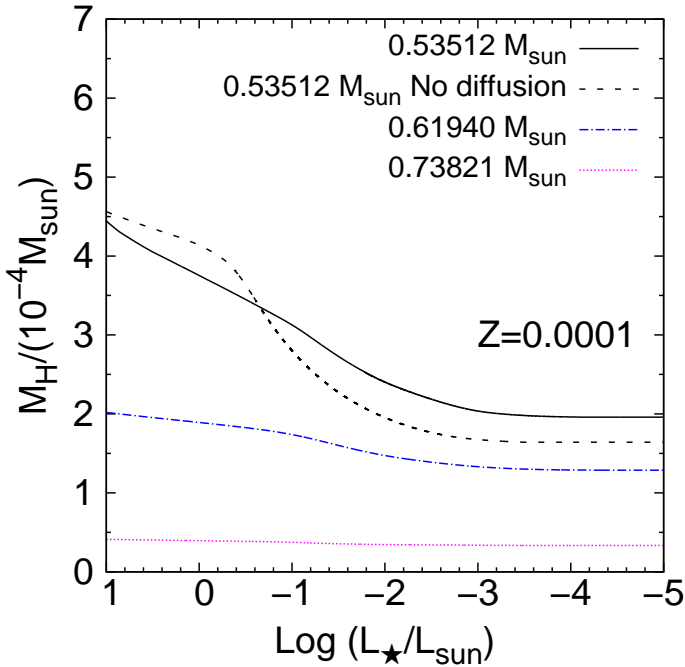}
\caption{Run of  the total H content  as a function of  luminosity for
  the 0.53512,  0.61940, and  $0.73821 \, M_{\sun}$  cooling sequences
  (solid, dot-dashed and dotted lines, respectively) Also shown is the
  H content  for the  $0.53512\, M_{\sun}$  white dwarf  sequence when
  element  difussion  processes  are   neglected  (dashed  line).  The
  metallicity of the progenitor is $Z=0.0001$.}
\label{Fig:MHvst}
\end{figure}

Figure~\ref{Fig:MHvst} shows the evolution of  the mass of H for three
selected  white dwarf  sequences  with $Z=0.0001$.   The  mass of  the
initial H layer decreases -- by almost a factor of two -- for a period
during which H burning supplies most  of the surface luminosity of the
white dwarf.   However, as soon as  the mass of the  H layer decreases
below a certain threshold, which depends  on the white dwarf mass, the
pressure at the bottom of the  envelope is unable to no longer support
nuclear  reactions, and  the  total  H mass  of  the  model reaches  a
constant value.  For comparison purposes, the run of the mass of H for
the $0.53512\,  M_{\sun}$ white dwarf sequence  when element difussion
processes are neglected is also shown  in the figure (dashed line). In
the absence of  element diffusion, a larger fraction of  the H content
of the star burnt at higher luminosities, as compared with the case in
which diffusion is allowed to operate.

\begin{figure*}
\centering
\includegraphics[clip,width=\textwidth]{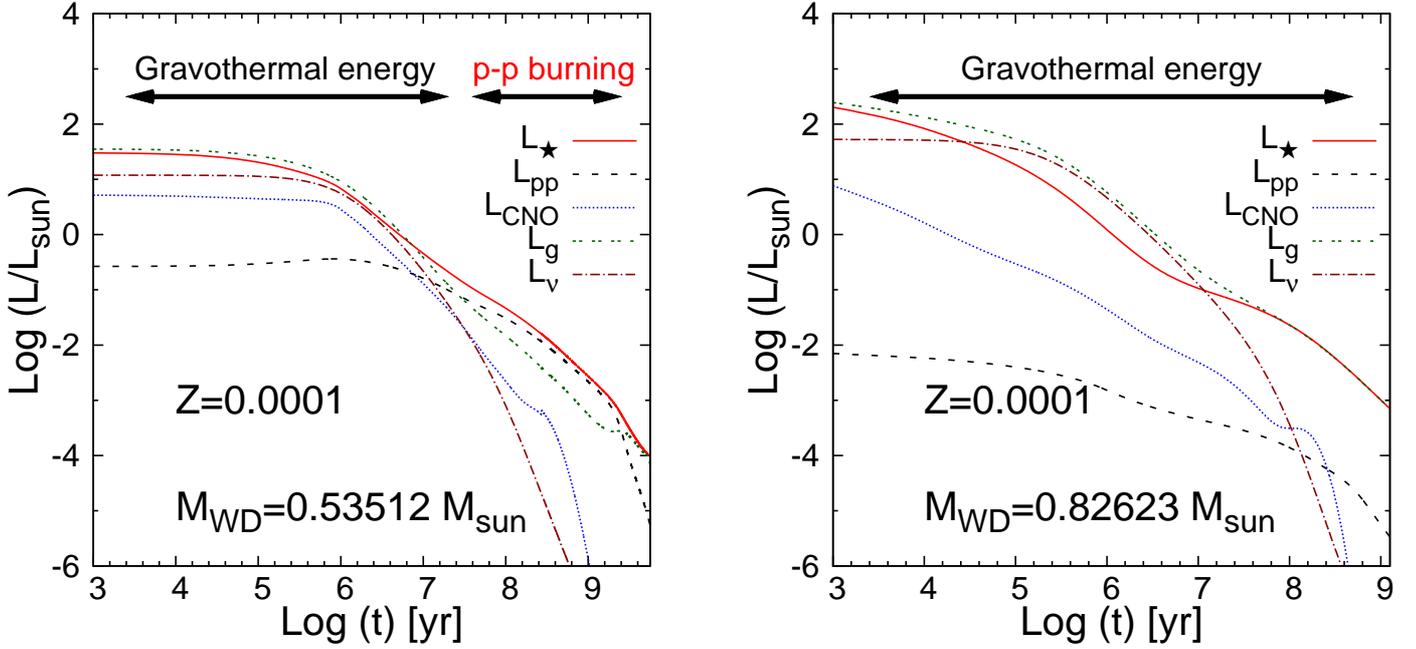}
\caption {Time dependence of  the various energy sources and sinks for
  our  $0.53512\,  M_{\sun}$  and  $0.82623\,  M_{\sun}$  white  dwarf
  sequences resulting from progenitors with $Z=0.0001$. It is depicted
  the surface  luminosity, $L_{\star}$ (solid line),  the luminosities
  arising from the proton-proton chain $L_{\rm pp}$ (dashed line), and
  from the CNO  bi-cycle $L_{\rm CNO}$ (dotted line),  the energy loss
  by   neutrino  emission   $L_{\nu}$  (dot   dashed  line)   and  the
  gravothermal (compression  plus thermal) energy release  $L_{\rm g}$
  (long dotted  line). The domain  of the  main energy source  is also
  indicated.  Time is measured in years since the moment when the star
  reaches the  maximum effective  temperature. The  progenitors masses
  are 0.85 and $2.5\, M_{\sun}$ respectively.}
\label{Fig:contribucionesaL}
\end{figure*}

In Fig.~\ref{Fig:contribucionesaL} we show  the time dependence of the
various energy sources and sinks during  the white dwarf cooling phase
for the  $0.53512\, M_{\sun}$ and the  $0.82623\, M_{\sun}$ sequences.
These  sequences  result,  respectively, from  progenitor  stars  with
masses 0.85 and  $2.5\, M_{\sun}$, and a  metallicity $Z=0.0001$.  For
the more massive  sequence, except for the short-lived  stages at very
high  luminosities  (not  shown  in   the  figure),  H  burning  never
represents an appreciable  source of energy.  This  is quite different
from the situation  for the $0.53512\, M_{\sun}$  sequence.  Here, the
H-burning  energy  release  resulting  from  the  proton-proton  chain
becomes  the main  source  of energy  over large  periods  of time  at
advanced ages, providing until $\log t \sim 9.4$ a contribution to the
surface luminosity that is much larger  than that given by the release
of gravothermal  energy.  The impact  of proton-proton burning  on the
resulting cooling curve  (solid line) is apparent.   Indeed, note that
towards the end  of proton-proton burning phase, there is  a change of
slope  in the  cooling curve,  until  crystallization sets  in at  the
center of  the white  dwarf and  the cooling  process slows  down (for
surface luminosities smaller than  $\log(L/L_{\sun})\sim -4$). This is
a result of the release of latent heat and gravitational energy caused
by  carbon-oxygen phase  separation.   In passing,  we  note that  the
contribution  of  the  CNO  bi-cycle  to  the  surface  luminosity  is
restricted   to    the   first   $10^6$~yr   of    the   white   dwarf
evolution. However, except at very high luminosities, its contribution
is always much smaller than  the release of gravothermal energy, which
drives evolution during these first stages.

\begin{figure}
\centering
\includegraphics[clip,width=\columnwidth]{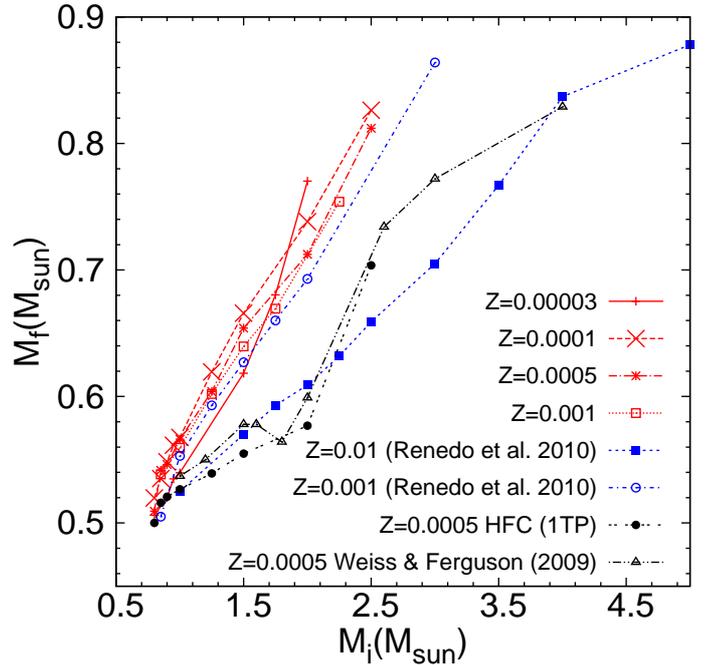}
\caption{Theoretical  initial-to-final mass  relations resulting  from
  our   sequences.    In  addition,   we   display   the  results   of
  \cite{2010ApJ...717..183R}   for  progenitors   with  $Z=0.01$   and
  $Z=0.001$, the  mass of the H  free core at the  first thermal pulse
  for   our  $Z=0.0005$   sequences,  and   the  initial-to-final mass
  relationship    of   \cite{2009A&A...508.1343W}    for   the    same
  metallicity.}
\label{Fig:MiMf}
\end{figure}

As mentioned  before, all the  white dwarf sequences computed  in this
work have been obtained  following the progenitor evolution throughout
the entire thermally-pulsing  AGB phase.  That is, we  have not forced
our models  to abandon early  this phase in  order to get  white dwarf
structures.  Given the fact that the mass loss adopted is just a rough
extrapolation  of   what  is   known  at  higher   metallicities,  the
initial-to-final  mass   relation  offers  a  good   testbed  for  our
sequences.   Then,  we   believe  it   is  worthwhile   to  show   the
initial-to-final mass  relation resulting  from our  full evolutionary
sequences. This  is done  in Fig.~\ref{Fig:MiMf}, where  our resulting
white  dwarf masses  as a  function of  the initial  mass at  ZAMS are
compared   with   the   results  of   \cite{2010ApJ...717..183R}   for
progenitors  with $Z=0.01$,  and $Z=0.001$,  and with  the final  mass
obtained by \cite{2009A&A...508.1343W} for $Z=0.0005$. In addition, we
include the mass  of the H-free core  at the first thermal  pulse as a
function  of  the  initial  mass   of  the  progenitor  for  the  case
$Z=0.0005$.   It should  be mentioned  that \cite{2009A&A...508.1343W}
consider overshooting  during the  thermally-pulsing AGB  phase, which
considerably  reduces the  growth of  the H-free  core. Thus,  in this
case, the  final mass is  not expected to  be very different  from the
mass  of the  H-free  core  at the  first  thermal  pulse.  With  this
consideration in mind it is worth noting the good agrement between our
mass of the H-free core at the first thermal pulse with the final mass
found by  \cite{2009A&A...508.1343W}.  By comparing our  two sequences
with $Z=0.0005$  it is clear that,  in the absence of  overshooting in
the thermally-pulsing AGB phase, as assumed  in our study, the mass of
the  H-free  core  grows  considerably  during  the  thermally-pulsing
AGB. Hence, in this case, the final mass will be strongly dependent on
the  assumed mass-loss  rates. This  explains the  steep slope  of the
initial-final mass relationship we find for the sequences presented in
this  study,  since smaller  mass-loss  rates  are expected  at  lower
metallicities. Nevertheless, it should  be emphasized that theoretical
initial-to-final mass  relationships can not be  accurately predicted,
as   they   depend    on   two   badly   known    processes   in   the
thermally-pulsing-AGB phase, i.e., winds and the occurrence of a third
dredge up episode.

\subsection{The importance of mass loss and overshooting}
\label{mlossov}

\begin{figure}
\centering
\includegraphics[clip,width=\columnwidth]{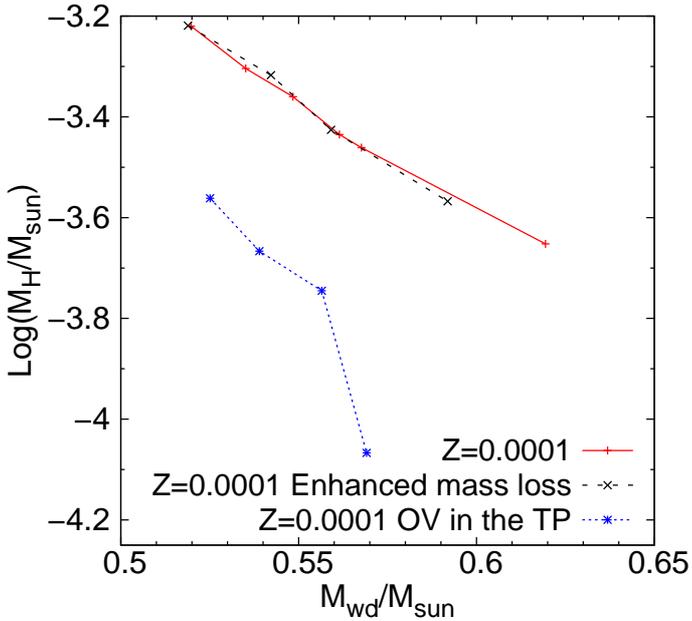}
\caption{Logarithm of  the total  mass of  H (in  solar units)  at the
  point  of  maximum effective  temperature  of  the cooling  sequence
  versus  the white  dwarf mass.  The  dotted (solid)  line shows  the
  predictions from our $Z=0.0001$ progenitors when overshooting during
  the thermally-pulsing AGB is considered (neglected). In addition, we
  show the results  for the case of an enhanced  mass loss rate during
  the thermally-pulsing  AGB and  the following early  post-AGB phases
  (dashed line).}
\label{mdot}
\end{figure}

We have shown  that the mass of H  that is left on the  white dwarf at
the beginning of  its cooling branch is independent  of the occurrence
of mass loss during the planetary nebulae stage.  Here, we explore the
role played  by mass  loss during the  thermally-pulsing AGB  phase in
fixing  the  final  H  content  of the  white  dwarf,  and  hence  the
importance of residual H burning.  We concentrate on those white dwarf
masses for which we  find residual H burning to be  the main source of
white dwarf energy.  To this end, we re-computed the full evolution of
our 0.80, 0.90, 1.0, and $1.25\, M_{\sun}$ progenitors with $Z=0.0001$
from the first thermal  pulse on the AGB all the  way to the beginning
of the white dwarf cooling  track, by artificially increasing the mass
loss  rates  by a  factor  of  two. This  is  done  during the  entire
thermally-pulsing  AGB,  as well  as  during  the subsequent  post-AGB
stage.  The resulting mass of H at the beginning of the cooling branch
predicted   by   these   numerical   experiments   is   displayed   in
Fig.~\ref{mdot} with  dashed line.  These predictions are  compared to
the results obtained whith our standard $Z=0.0001$ progenitors with no
enhanced mass loss.  As expected, increasing the mass-loss rate yields
smaller white  dwarf masses.  But note  that for  a given  white dwarf
mass, the  final mass of  H is the  same, irrespective of  the adopted
mass-loss rate. In  fact, the agreement in the relation  H mass versus
white  dwarf mass  is  surprisingly good.   This important  experiment
shows that the H  content with which a white dwarf  of a given stellar
mass enters its cooling track is  independent of the adopted mass loss
rates during progenitor evolution.  Hence, our predictions that stable
nuclear burning is the main source of energy over most of evolution of
low-mass  white dwarfs  coming  from  low-metallicity progenitors  are
valid independently  of the rate  at which  mass is lost  during prior
evolution.

\begin{figure}
\centering
\includegraphics[clip,width=\columnwidth]{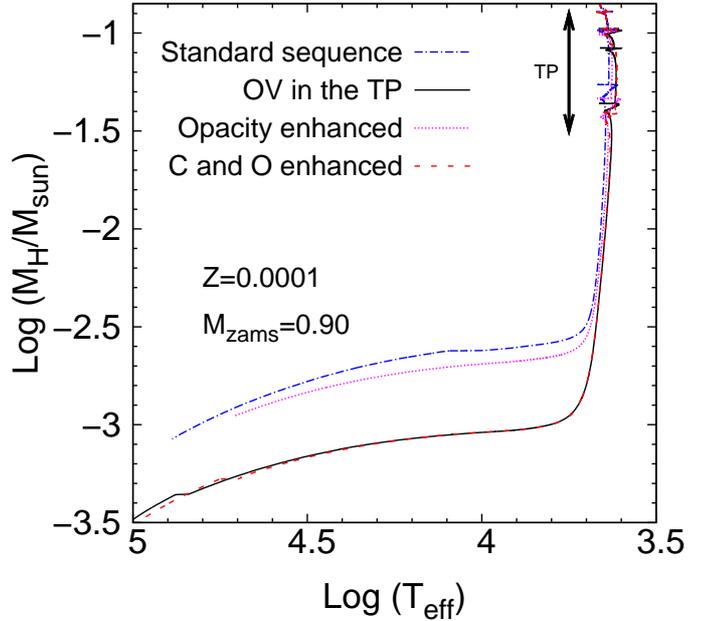}
\caption{Evolution of the residual H content (in solar units) in terms
  of  the effective  temperature for  our $0.9\,  M_{\sun}$ progenitor
  with   $Z=0.0001$   during   the  thermally-pulsing   and   post-AGB
  phases.  Our  standard  sequence   is  depicted  with  a  dot-dashed
  line.    The   sequence    including    overshooting   during    the
  thermally-pulsing  AGB  phase  is  shown  with  a  solid  line.  Two
  additional artificial  sequences with no overshooting  are included:
  one  in which  the carbon  and oxygen  enrichment expected  from the
  third  dredge-up   is  only  considered  in   the  envelope  opacity
  computation (dotted line),  and another one in  which the abundances
  of carbon and oxygen have  been altered according to the predictions
  of the third dredge-up, see text for details.}
\label{mteff}
\end{figure}

On  the  contrary,  and  in  view  of  our  previous  discussion,  the
occurrence of  third dredge-up during the  thermally-pulsing AGB phase
is expected  to reduce  the final  H content, and  thus the  impact of
residual nuclear burning. To be more quantitative, we have re-computed
some of our sequences but now allowing overshooting from the beginning
of the  thermally-pulsing AGB  phase.  It  is well  known that  at low
metallicities overshooting favors the  occurrence of a third dredge-up
episode      also     for      very      small     stellar      masses
\citep{2009A&A...508.1343W}. To  this end, we follow  the evolution of
our  0.85,  0.90, 1.0  and  $1.25\,  M_{\sun}$ progenitor  stars  with
$Z=0.0001$   from  the   first  thermal   pulse  including   diffusive
overshooting with the overshooting parameter  set to $f=0.015$ at each
convective     boundary      --     see     \cite{2010ApJ...717..183R,
2009A&A...508.1343W} for details.  Now, all these sequences experience
appreciable third  dredge-up, with  the result that  the C/O  ratio is
strongly increased  (see Fig.~\ref{Fig:CO}).  Note that  the C/O ratio
reaches values of  about 3--6, in good agreement  with the predictions
of \cite{2009A&A...508.1343W} for  $Z=0.0005$ progenitors.  The carbon
enrichment of the envelope by more than three orders of magnitude as a
result of a third dredge-up episode eventually leads to a more intense
CNO  burning,  which strongly  reduces  the  final  H content  of  the
resulting white dwarf, as can be appreciated in Fig.~\ref{mdot}.  Note
that a reduction by more than a factor of two in the mass of the final
H layer  is expected  if third dredge-up  takes place.   This strongly
reduces  the impact  of  residual  H burning  during  the white  dwarf
regime.  We conclude that the occurrence of third dredge-up during the
thermally-pulsing AGB evolution of low-mass stars (due to for instance
to the  occurrence of  overshooting) prevents H  burning from  being a
relevant   energy  source   in  cool   white  dwarfs   resulting  from
low-metallicity progenitors. In this  sense, since the third dredge-up
is strongly reduced  in the sequences we presented in  this paper (due
to our  neglect of  overshooting), the  final H  content of  our white
dwarf sequences  and the  consequent impact of  residual H  burning on
white dwarf cooling times should be considered as upper limits.

In  order  to   obtain  a  better  understanding  of   the  impact  of
overshooting on the final H content, we have re-computed the evolution
of the $0.9\, M_{\sun}$ progenitor star with $Z=0.0001$ from the first
thermal pulse,  this time neglecting  overshooting (and hence  with no
occurrence  of dredge-up)  but  artificially taking  into account  the
carbon and oxygen enrichment expected from the third dredge-up only in
the  envelope opacity.   The  result  is shown  as  a  dotted line  in
Fig.~\ref{mteff}, in which  the resulting evolution of  the residual H
content  (in  solar units)  is  depicted  in  terms of  the  effective
temperature  during the  thermally-pulsing  and  post-AGB phases.  Our
standard sequence with no overshooting  is depicted using a dot-dashed
line,  together with  the prediction  of the  same sequence  including
overshooting during the thermally-pulsing  AGB phase (solid line).  It
is clear from this figure that most  of the reduction of the H content
that characterizes the  sequences with overshooting is  not the result
of  an   opacity  effect.   Another  artificial   sequence  neglecting
overshooting  was computed  for  which the  abundances  of carbon  and
oxygen in the entire model envelope have been altered according to the
predictions expected from the third dredge-up (dashed line).  In doing
this,  we  neglect  any  other  consequence  of  overshooting  on  the
evolution.  Clearly,  it is  the chemical  enrichment of  the envelope
that is  responsible for the  thinner H  envelopes, not as  an induced
opacity effect, but as an increase in the H burning via CNO.

\section{Summary and conclusions}
\label{conclusions}

In view  of the  recent finding  that residual  H burning  in low-mass
white dwarfs resulting  from $Z=0.0001$ progenitors could  be the main
energy      source     over      most      of     their      evolution
\citep{2013ApJ...775L..22M},  in this  paper  we  presented new  white
dwarf   evolutionary  sequences   resulting   from  progenitor   stars
characterized by  a wide range  of low-metallicity values.   These new
sequences  are  thus  appropriate  for the  study  of  low-metallicity
stellar populations.   We computed the  full evolution of  white dwarf
stars by  consistently evolving the  progenitor stars through  all the
stellar phases,  including the  ZAMS, the red  giant branch,  the core
helium flash (when occurred), the  stable core helium burning, the AGB
phase  and the  entire  thermally-pulsing and  post-AGB phases.   Four
progenitor  metallicities  have  been considered:  $Z=0.001$,  0.0005,
0.0001 and 0.00003.  To our knowledge,  this is the first set of fully
evolutionary   calculations    of   white   dwarfs    resulting   from
low-metallicity  progenitors covering  the relevant  range of  initial
main  sequence and,  correspondingly, white  dwarf masses.  A relevant
aspect of our sequences is that the masses of the H-rich envelopes and
of  the helium  shells were  obtained from  evolutionary calculations,
instead of  using typical  values and  artificial initial  white dwarf
models.   In  computing  this  new   set  of  white  dwarf  sequences,
overshooting   was  disregarded   during  the   thermally-pulsing  AGB
phase. Consequently,  third dredge-up is  much less efficient  than in
sequences  that include  overshooting  at  all convective  boundaries.
Independently of the fact that  evidence is not completely clear about
the  possible  occurrence  of   appreciable  overshooting  during  the
thermally-pulsing AGB phase (even more in the low-metallicity regime),
our reason  for this  choice is  that we  are interested  in providing
white dwarf sequences for which the residual H burning has its maximum
impact on white dwarf cooling times (see below).

We found that in the absence  of third dredge-up episodes, most of the
evolution  of low-mass  white  dwarfs  resulting from  low-metallicity
progenitors in  the range $0.00003  \la Z  \la 0.001$ is  dominated by
stable H  burning.  Specifically, we  find that for white  dwarfs with
$M\la 0.6  \,M_{\sun}$ energy  released by  H burning  constitutes the
main   energy  source   at   intermediate   luminosities  ($-3.2   \la
\log(L/L_{\sun})\la -1$), thus resulting  in substantial delays in the
cooling  times.   At  intermediate luminosities,  residual  H  burning
increases the cooling  times by more than  a factor of 2,  and even at
$\log(L/L_{\sun})\sim  -4$,  such  delays   amounts  to  $20-40\%$  in
low-mass white dwarfs.

Our finding that  stable H burning is a main  energy source that could
be sustained over a long period  of evolution in low-mass white dwarfs
that  result from  low-metallicity progenitors  opens the  possibility
that this  stable burning  could impact  on the  pulsational stability
properties of these white dwarfs.  The existence of low-order $g$-mode
destabilized by  the $\varepsilon-$mechanism  in hot  H-rich pre-white
dwarf models  has recently been  reported \citep{2014arXiv1405.4568M}.
The  assessment  of  fully  nonadiabatic pulsation  properties  and  a
stability analysis of our models  constitute an important issue, since
this   could    provide   for    the   first   time    evidence   that
$\varepsilon$-destabilized  modes could  be  expected in  evolutionary
models   of  cool,   standard  carbon-oxygen   white  dwarfs   --  see
\cite{2014arXiv1408.6724C}  for  a  recent   report  of  evidence  of
$\varepsilon$-destabilized modes in helium-core white dwarfs. However,
this  issue is  beyond the  scope of  the present  paper, and  will be
explored in a near future.

We have  also shown  that our prediction  that stable  nuclear burning
might be the main source of  energy over most of evolution of low-mass
white dwarfs resulting from  low-metallicity progenitors remains valid
independently of  the rate at  which stellar  mass is lost  during the
thermally-pulsing and post-AGB phases, and during the planetary nebula
stage.  However,  we find that  the mass of  H of the  resulting white
dwarfs becomes sustantially  smaller (by a factor of two)  as a result
of the occurrence of third  dredge-up during the thermally-pulsing AGB
phase.   In this  sense, we  find that  the carbon  enrichment of  the
progenitor  envelope resulting  from  the third  dredge-up prevents  H
burning  from being  a relevant  energy  source in  cool white  dwarfs
resulting from low-metallicity progenitors.

According to our simulations,  white dwarfs populating low-metallicity
globular  clusters could  be characterized  by significant  residual H
burning, so  as to leave  some signature  on the observed  white dwarf
luminosity  function  of  such   clusters.   NGC~6397,  a  metal  poor
([Fe/H]$=-2.1$),  old globular  cluster,  whose well-determined  white
dwarf luminosity  function has  been proposed as  a tool  to constrain
white  dwarf   physics  \citep{2009ApJ...693L...6W}   is  particularly
relevant in this  context.  In particular, this cluster  could be used
as a  testbed for the  occurrence of  stable nuclear burning  in their
white dwarfs,  which, in view  of the results  of our study,  could be
used in turn  to constrain the efficiency of  third dredge-up episodes
and  the ensuing  carbon enrichment  during the  thermally-pulsing AGB
phase of low-metallicity progenitors.

\begin{acknowledgements}
We  thank Aldo  Serenelli for  helping  us to  improve some  numerical
aspects of {\tt LPCODE} evolutionary code. We acknowledge the comments
and suggestions of our referee  which improved the original version of
this  paper.   M3B  is  supported by  a  fellowship  for  postdoctoral
researchers from the Alexander von  Humboldt Foundation.  Part of this
work was supported by AGENCIA  through the Programa de Modernizaci\'on
Tecnol\'gica BID  1728/OC-AR, by  the PIP 112-200801-00940  grant from
CONICET, by MCINN grant AYA2011-23102, and by the European Union FEDER
funds.  This research has made use of NASA Astrophysics Data System.
\end{acknowledgements}

\bibliographystyle{aa} 

\end{document}